\documentclass[a4paper]{article}

\usepackage{INTERSPEECH2020}
\usepackage{multirow}
\usepackage{caption}
\usepackage{subcaption}
\usepackage{tikz}
\usepackage{pgfplots}
\usepgfplotslibrary{groupplots}

\title{Improving Cross-Lingual Transfer Learning for End-to-End Speech Recognition with Speech Translation}
\name{Changhan Wang, Juan Pino, Jiatao Gu}
\address{Facebook AI, USA}
\email{\{changhan, juancarabina, jgu\}@fb.com}

\begin{document}

\maketitle
\begin{abstract}
Transfer learning from high-resource languages is known to be an efficient way to improve end-to-end automatic speech recognition (ASR) for low-resource languages. Pre-trained or jointly trained encoder-decoder models, however, do not share the language modeling (decoder) for the same language, which is likely to be inefficient for distant target languages. We introduce speech-to-text translation (ST) as an auxiliary task to incorporate additional knowledge of the target language and enable transferring from that target language. Specifically, we first translate high-resource ASR transcripts into a target low-resource language, with which a ST model is trained. Both ST and target ASR share the same attention-based encoder-decoder architecture and vocabulary. The former task then provides a fully pre-trained model for the latter, bringing up to 24.6\% word error rate (WER) reduction to the baseline (direct transfer from high-resource ASR). We show that training ST with human translations is not necessary. ST trained with machine translation (MT) pseudo-labels brings consistent gains. It can even outperform those using human labels when transferred to target ASR by leveraging only 500K MT examples. Even with pseudo-labels from low-resource MT (200K examples), ST-enhanced transfer brings up to 8.9\% WER reduction to direct transfer.
\end{abstract}


\section{Introduction}

The attention-based encoder-decoder model paradigm~\cite{sutskever2014sequence,bahdanau2014neural} has recently witnessed rapidly increased applications in end-to-end automatic speech recognition (ASR). It provides a generic framework for speech-to-text generation tasks, and achieves state-of-the-art performance on ASR~\cite{chiu2017stateoftheart,park2019specaugment,synnaeve2019end} as an alternative to CTC (Connectionist temporal classification) models~\cite{graves2006connectionist}. The recent surge of end-to-end speech-to-text translation (ST) studies~\cite{berard2016listen,duong2016attentional,weiss2017sequence,vila2018end,berard2018end} is also due to the application of attention-based encoder-decoder models. And very recent works~\cite{anastasopoulos2018tied,di2019one,liu2019synchronous} have demonstrated the possibility of combining the two related tasks, ASR and ST, under the same encoder-decoder architecture to achieve better performance. When targeting at ST only, transfer learning from ASR~\cite{bansal2018pre,di2019one} is helpful to warm-starting acoustic modeling (encoder) and enabling ST model training to focus more on learning language modeling and alignment (decoder).

In this paper, we study how to utilize ST to improve cross-lingual transfer learning for ASR. Transfer learning from high-resource languages~\cite{dalmia2018sequence,yi2018adversarial,cho2018multilingual,zhou2018multilingual} is known to be an efficient way to improve end-to-end ASR for low-resource languages. Pre-trained or jointly trained encoder-decoder models, however, do not share the language modeling (decoder) for the same language, which is likely to be inefficient for distant target languages. We introduce ST as an auxiliary task to incorporate additional knowledge of the target language and enable transferring from that target language. Unlike previous ideas for leveraging translation data~\cite{anastasopoulos2018leveraging,hayashi2018back,Wiesner2019}, our approach does not require any modification to the ASR model architecture. It leverages ST data instead of text-to-text translation data for ST training, which avoids speech-to-text modality adaption in the encoder. Moreover, we train ST with machine translation (MT) pseudo-labels on high-resource ASR transcripts, which overcomes the shortage of real ST data and consistently brings gains to the transfer learning. MT pseudo-labeling also simplifies ST model training (knowledge distilled data) and allows beam-searching diverse labels to alleviate overfitting.

\begin{figure}[t]
    \centering
    \includegraphics[width=\linewidth]{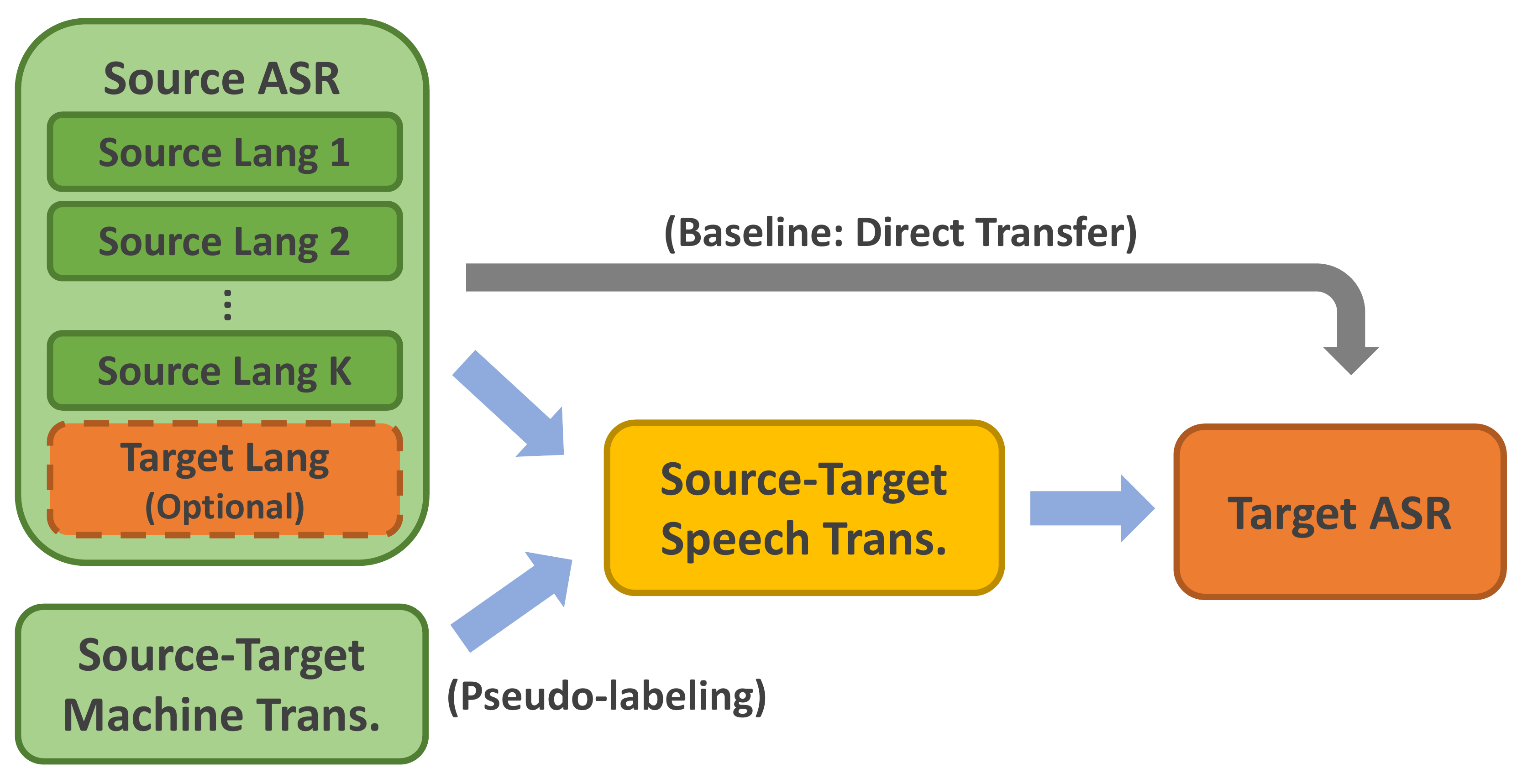}
    \caption{An overview of proposed cross-lingual transfer learning pipeline. The color reflects data availability/quality.}
    \label{fig:overview}
\end{figure}

\section{Methods}

\subsection{Attention-Based Encoder-Decoder Architecture}
Our ASR and ST models share the same BLSTM-based encoder-decoder architecture~\cite{berard2018end} with attention mechanism, which is similar to the LAS architecture~\cite{chan2016listen,chiu2017stateoftheart,park2019specaugment}. Specifically, on the encoder side, audio features $\textbf{x}\in\mathbb{R}^{T\times d_0}$ are first fed into a two-layer DNN with $tanh$ activations and hidden sizes $d_1$ and $d_2$. Then two 2D convolutional layers with kernel size $3$x$3$ and stride $2$x$2$ are applied to reduce the sequence length to $\frac{T}{4}$. Both convolutional layers have 16 output channels and project the features to $4d_2$ dimensions after flattening. Finally, the features are passed to a stack of three bidirectional LSTM layers of hidden size $d_3$ to form encoder output states $\textbf{h}\in\mathbb{R}^{T\times 2d_3}$. For the decoder side, a stack of two LSTM layers with hidden size $2d_3$ and additive attention~\cite{bahdanau2014neural} is applied, followed by a linear projection to size $d_o$.

For MT, we use one of Transformer \emph{base} with 3 encoder/decoder layers, Transformer \emph{base} and Transformer \emph{big} models~\cite{vaswani2017attention} (with original training hyper-parameters) depending on the MT dataset size.

\subsection{Speech Translation Trained with Pseudo-Labels}
Word-level or sequence-level knowledge distillation (KD) is helpful to MT~\cite{kim2016sequence} and ST~\cite{liu2019end} model training, because it reduces noise and simplifies data distribution in the training set. Training end-to-end ST models is known to be difficult, given the fact that it needs to learn acoustic modeling, language modeling and alignment at the same time. When the training data distribution is complex, end-to-end ST models are likely to fit the data worse than cascading ASR and MT models. Moreover, ST labels are more expensive to obtain than ASR or MT ones. Existing ST corpora are strongly limited by size and language coverage, making ST model training even more difficult. To overcome the shortage of real data, we propose to pseudo-label ASR corpora with MT and train ST on the resulting datasets. This provides larger scale training data as well as more diversity (via different MT models and beam search) at little cost. Both are useful for alleviating overfitting. Moreover, training ST models with MT pseudo-labels can be viewed as a sequence-level KD process. Although potentially inaccurate pseudo-labels can hurt model training, pseudo-labels are easier to be fitted. This compensates its gap towards real labels, which are likely more difficult to learn. In our experiments, we show that ST models trained with pseudo-labels can even outperform those using real labels when transferred to the target ASR.

\subsection{Pre-training ASR on Speech Translation}

Instead of pretraining target (low-resource) ASR directly on (multilingual) source (high-resource) ASR, we pretrain target ASR on source-to-target ST. The latter is pretrained on source ASR and leverages MT pseudo-labels on source ASR data for training. Figure~\ref{fig:overview} provides an overview of our proposed transfer learning pipeline: $\text{ASR}_{\text{Source}}\rightarrow\text{ST}_{\text{Source-Target}}\rightarrow\text{ASR}_{\text{Target}}$. Our intuition is that this two-step approach helps to decouple transfer of language modeling (decoder) and acoustic modeling (encoder) to make transfer learning smoother and more effective. Moreover, the ST model leverages additional data (MT pseudo-labels) for the target language and hence is likely to model the target language better. We use the same model architecture for ASR and ST, so that they can be easily transferred between each other: $\text{ASR}_{\text{Source}}\rightarrow\text{ST}_{\text{Source-Target}}$\footnote{ Full model transfer excluding the embedding and softmax layers.} and $\text{ST}_{\text{Source-Target}}\rightarrow\text{ASR}_{\text{Target}}$. Pretraining ST with ASR warm-starts acoustic modeling so that ST training can be more focused on learning language modeling and alignment. We may simplify the transfer learning pipeline by training ST from scratch. This still outperforms direct $\text{ASR}_{\text{Source}}\rightarrow\text{ASR}_{\text{Target}}$ transfer in most of the cases as shown in our experiments. To another extreme, we may pre-train ST jointly on source+target ASR to warm-start from a better acoustic model: $\text{ASR}_{\text{Source+Target}}\rightarrow\text{ST}_{\text{Source-Target}}$.

\section{Experiments}

\begin{table}[t]
  \caption{Source and target ASR data.}
  \label{tab:asr_data}
  \centering
  \begin{tabular}{c|ccc}
    \toprule
    & Dataset & Train & Speakers \\
    \midrule
    \multicolumn{4}{c}{Source ASR} \\
    \midrule
    CV & Common Voice: English & 477h & 15.2k \\
    CV$_{Fr}$ & Common Voice: French & 264h & 1.8k \\
    LS & Librispeech & 960h & 2.3k \\
    MC & MuST-C: En-Nl & 422h & 2.2k \\
    \midrule
    \multicolumn{4}{c}{Target ASR} \\
    \midrule
    Vi & IARPA Babel 107b-v0.7 & 96h & 0.6k \\
     Ht & IARPA Babel 201b-v0.2b & 70h & 0.3K \\
    Pt & Common Voice v4 & 10h & 2 \\
    Zh-CN & Common Voice v4 & 10h & 22 \\
    Nl & Common Voice v4 & 7h & 78 \\
    Mn & Common Voice v4 & 3h & 4 \\
    \bottomrule
  \end{tabular}
  \vspace{-3mm}
\end{table}

\begin{table}[t]
    \caption{MT data and Transformer models.}
    \small
    \centering
    \begin{tabular}{c|ccc}
    \toprule
    & Dataset & En/Fr Sent. & Model \\
    \midrule
    Vi & OpenSubtitles & 4M/3M & Base \\
    Ht & JW300 & 220K/220K & Base 3+3 \\
    Pt & OpenSubtitles & 33M/23M & Big \\
    Zh & MultiUN & 10M/10M & Big \\
    Nl & OpenSubtitles & 37M/25M & Big \\
    Mn & JW300+GNOME+QED & 210K/203K & Base 3+3 \\
    Nl$_{W}$ & WikiMatrix & 511K/- & Base 3+3 \\
    Nl$_{S}$ & OpenSubtitles & 37M/- & Base 3+3 \\
    Nl$_{M}$ & OpenSubtitles & 37M/- & Base \\
    \bottomrule
    \end{tabular}
    \label{tab:mt_data}
    \vspace{-3mm}
\end{table}

\subsection{Data}
For English and English+French ASR, we use Librispeech~\cite{panayotov2015librispeech} and Common Voice~\cite{ardila2019common} (v4, 2019-12-10 release).\footnote{The original dataset splits contain only one sample per sentence. We instead use extended splits~\cite{wang-etal-2020-covost} to allow using all samples.} We also use the ASR data in MuST-C~\cite{di2019must} (En-Nl subset) for the analysis in section~\ref{section:different_pl}. For target ASR, we use Portuguese (Pt), Chinese (Zh-CN), Dutch (Nl) and Mongolian (Mn) from Common Voice v4$^2$ as well as Vietnamese (Vi) and Ht (Haitian) from IARPA Babel datasets (conversational telephone speech). Basic statistics of all used ASR corpora can be found in Table~\ref{tab:asr_data}. For MT, we use a variety of datasets indexed by OPUS~\cite{tiedemann2012parallel}, which are listed in Table~\ref{tab:mt_data}.

\subsection{Experimental Setup}
For all texts, we normalize their punctuation and tokenize them with sacreMoses\footnote{https://github.com/alvations/sacremoses}. For ASR and ST, we lowercase the texts (except for Babel). For ASR, we remove all punctuation markers except for apostrophes. We use character vocabularies for ASR and ST, and use BPE vocabularies~\cite{sennrich2015neural} for MT. We extract 80-channel log-mel filterbank features (windows with 25ms size and 10ms shift) using Kaldi~\cite{povey2011kaldi}, with per-utterance cepstral mean and variance normalization (CMVN) applied. We remove training samples having more than 3,000 frames or more than 512 characters for GPU memory efficiency.

The configuration of MT models can be found in Table~\ref{tab:mt_data}. For ASR and ST models, we set $d_1=256$, $d_2=128$, $d_3=512$ and $d_o=128$. We adopt SpecAugment~\cite{park2019specaugment} (LB policy without time warping) to alleviate overfitting. All models are implemented in Fairseq~\cite{ott2019fairseq}. We use a beam size of 5 for decoding. We average the last 5 checkpoints for ASR and ST, and average the last 2 checkpoints for MT. For MT and ST, we report case-insensitive tokenized BLEU~\cite{papineni2002bleu} using sacreBLEU~\cite{post-2018-call}. For ASR, we report character error rate (CER) on Chinese (no word segmentation) and word error rate (WER) on the other languages using VizSeq~\cite{Wang_2019}.

\begin{table*}[t]
    \caption{Test WER (relative reduction in parentheses) for cross-lingual transfer from English and from English+French}
    \centering
    \begin{tabular}{lr|cc|cccc}
    \toprule
    & & Vi & Ht & Pt & Zh-CN & Nl & Mn \\
    \midrule
    \multicolumn{2}{c|}{Baseline} & 57.2 & 66.1 & 62.3 & 90.3 & 96.5 & 109.7 \\
    \midrule
    \multicolumn{8}{c}{From English} \\
    \midrule
    \multirow{5}{*}{CV} & Src ASR & 53.7 & 60.7 & 40.9 & 41.3 & 44.2 & 67.7 \\
    & + ST & 52.5 (-2.2\%) & 59.3 (-2.3\%) & 33.7 (-17.6\%) & 35.3 (-14.5\%) & 42.0 (-5.0\%) & 64.1 (-5.3\%)\\
    & Src+Tgt ASR & 51.6 & 58.1 & 34.7 & 37.0 & 42.5 & 63.0 \\
    & + ST & 51.2 (-0.8\%) & 57.2 (-1.5\%) & 31.2 (-10.1\%) & 35.2 (-4.9\%) & 40.4 (-4.9\%) & 62.3 (-1.1\%) \\
    \midrule
    \multirow{4}{*}{CV+LS} & Src ASR & 54.7 & 59.9 & 41.3 & 40.0 & 42.2 & 66.1 \\
    & + ST & 52.9 (-3.3\%) & 57.4 (-4.2\%) & 31.8 (-23.0\%) & 35.7 (-4.2\%) & 37.9 (-10.2\%) & 60.2 (-8.9\%) \\
    & Src+Tgt ASR & 52.7 & 57.8 & 34.4 & 36.4 & 41.7 & 67.9 \\
    & + ST & 52.2 (-0.9\%) & 57.2 (-1.0\%) & 31.2 (-9.3\%) & 35.5 (-2.5\%) & 38.8 (-7.0\%) & 62.5 (-8.0\%) \\
    \midrule
    \multicolumn{8}{c}{From English+French} \\
    \midrule
    \multirow{4}{*}{CV+CV$_{Fr}$} & Src ASR & 54.5 & 59.4 & 39.5 & 39.2 & 43.0 & 67.7 \\
    & + ST & 51.7 (-5.1\%) & 57.8 (-2.7\%) & 29.8 (-24.6\%) & 33.6 (-14.3\%) & 38.4 (-10.7\%) & 62.1 (-8.3\%) \\
    & Src+Tgt ASR & 52.9 & 57.1 & 31.7 & 36.4 & 40.7 & 62.4 \\
    & + ST & 52.0 (-1.7\%) & 55.7 (-2.5\%) & 28.6 (-9.8\%) & 32.9 (-9.6\%) & 38.3 (-5.9\%) & 59.6 (-4.5\%) \\
    \bottomrule
    \end{tabular}
    \label{tab:lores_asr_wer}
\end{table*}

\begin{table}[t]
    \centering
    \caption{Test WER for source ASR}
    \begin{tabular}{c|ccc|cc}
    \toprule
    & CV & +LS & +CV$_{Fr}$ & MC & +CV \\
    \midrule
    En/Fr & 25.4/- & 16.7/- & 23.4/20.1 & 19.6/- & 18.6/- \\
    \bottomrule
    \end{tabular}
    \label{tab:en_asr_wer}
    \vspace{-3mm}
\end{table}

\subsection{Results}
\subsubsection{Cross-Lingual Transfer via ST}

We examine two settings for high-resource source ASR: monolingual (English) and multilingual (English and French). The test WER of source ASR models can be found in Table~\ref{tab:en_asr_wer}. Both settings use the same low-resource targets from different language families: Indo-European (Portugese and Dutch), Sino-Tibetan (Chinese), Austro-Asiatic (Vietnamese), French Creole (Haitian) and Mongolic (Mongolian). We experiment with different transfer learning strategies: with or without ST as an intermediate step, and with or without target ASR during ASR pre-training. The results (test WER) are presented in Table~\ref{tab:lores_asr_wer}: $\text{ASR}_\text{Source}\rightarrow\text{ASR}_\text{Target}$ (``Src ASR"), $\text{ASR}_\text{Source}\rightarrow\text{ST}_\text{Source-Target}\rightarrow\text{ASR}_\text{Target}$ (the 2nd row ``+ ST"); $\text{ASR}_\text{Source+Target}\rightarrow\text{ASR}_\text{Target}$ (``Src+Tgt ASR"), $\text{ASR}_\text{Source+Target}\rightarrow\text{ST}_\text{Source-Target}\rightarrow\text{ASR}_\text{Target}$ (the 4th row "+ ST"). We see that for both monolingual and multilingual settings, ST pre-training consistently brings gains to the direct transfer baseline. On Portuguese (Pt) and Dutch (Nl), there is over 9.3\% and 4.9\% WER reduction in all ST-enhanced transfers, respectively. There is also 1.0\%-8.9\% WER reduction on Haitian (Ht) and Mongolian (Mn), where MT is also low-resource with only around 200K training examples available. When the source ASR has larger scale data (from CV to CV+LS), the gains brought by ST may be enlarged, for example, from $5.0\%$ reduction to $10.2\%$ reduction for Dutch and from $5.3\%$ reduction to $8.9\%$ reduction for Mongolian.

\subsubsection{MT Models for Pseudo-Labeling}
\label{section:different_pl}
\begin{table}[t]
    \centering
    \caption{Performance of different Dutch pseudo-labels}
    \begin{tabular}{c|cccccc}
    \toprule
    & \multicolumn{6}{c}{ST Label (NA for baseline transfer)} \\
    & NA & Nl$_W$ & Nl$_S$ & Nl$_M$ & Nl & Real \\
    \midrule
    MT & - & 24.8 & 34.0 & 34.1 & 35.6 & \textbf{100.0} \\
    \midrule
    ST & - & 18.9 & 23.7 & 23.9 & \textbf{24.0} & 23.9 \\
    +CV & - & 18.6 & \textbf{23.3} & 22.6 & 23.1 & - \\
    \midrule
    ASR & 44.7 & \textbf{42.4} & 43.1 & 43.2 & 43.9 & 43.9 \\
    +CV & 42.4 & \textbf{38.7} & 40.0 & 39.2 & \textbf{38.7} & - \\
    \bottomrule
    \end{tabular}
    \label{tab:nl_asr}
    \vspace{-3mm}
\end{table}

In order to better understand how different MT pseudo-labels may affect the performance of ST as well as downstream target ASR, we experiment with Dutch pseudo-labels from different MT models for ST training: Nl$_W$ and Nl$_S$ both use Transformer \emph{base} with 3 encoder/decoder layers but are trained on WikiMatrix (0.5M examples) and OpenSubtitles (37M examples), respectively; Nl$_S$, Nl$_M$ and Nl are all trained on OpenSubtitles but use Transformer \emph{base} with 3 encoder/decoder layers, Transformer \emph{base} and Transformer \emph{big}, respectively. We use MuST-C optionally with Common Voice ("+CV") as English source ASR for different data conditions. Results (MT and ST BLEU on MuST-C test set as well as Dutch ASR test WER) are available in Table~\ref{tab:nl_asr}. We notice that the ST model using Nl has almost the same ST BLEU as that using real labels, although Nl has only 35.6 MT BLEU. Real labels are more difficult to learned in ST (76\% BLEU drop from MT to ST compared to pseudo-labels' 33\%). Nl$_W$ and Nl$_S$ share the same architecture, while the latter is trained on a more noisy corpus. When using MC only, Nl$_W$ outperforms Nl$_S$ and smaller models on OpenSubtitles (Nl$_S$ and Nl$_M$) are helpful to suppressing noise. When more data (MC+CV) is available, Nl$_S$ performs as well as Nl$_W$ on downstream ASR and larger models on OpenSubtitles are better for transfer.

\subsubsection{Pseudo-Label Sampling and Filtering}

Pseudo-labels are from beam search decoding of MT models. There are up to $k$ predictions per example given beam size $k$. Instead of using only the best ones for ST model training, we explore using the $n$-best ones ($2\leq n\leq k$) to provide more diversity and alleviate overfitting. Specifically, in each epoch, training labels are uniformly sampled from the set of $n$-best candidates. Pseudo-labels from low-resource or out-of-domain MT models may have low quality on some of the examples. We optionally filter 10\% examples by confidence scores (length-normalized log likelihood) to reduce noisy labels. We experiment with Dutch (highest MT resource) and Mongolian (lowest MT resource) for different values of $n$ ($k=5$). It can be seen from Figure~\ref{fig:label_sampling} that $n$-best pseudo-labels lead to lower dev WER in most of the cases and filtering helps significantly when MT is low-resource (Mongolian).

\subsubsection{Effectiveness of ST Pre-training}
We introduce ST to the pipeline with the idea of bringing pre-trained models closer to the target ones. In other words, we expect the $\text{ST}_\text{Source-Target}\rightarrow\text{ASR}_\text{Target}$ transfer to be faster than the $\text{ASR}_{\text{Source}}\rightarrow\text{ASR}_\text{Target}$ transfer. We examine the training accuracy curves for Vietnamese (highest resource) and Mongolian (lowest resource) to verify our hypothesis (see Figure~\ref{fig:acc_curve}). We observe that the ST-enhanced transfer (``w/ ST") has substantially higher starting points (60 to 53 and 70 to 36) and keeps leading with a substantial gap throughout the training process.

\subsubsection{ST without ASR Pre-training}
Instead of using ST as an intermediate step during transfer, we can also train ST from scratch to simplify the transfer pipeline. We experiment with CV+CV$_{Fr}$, whose results can be found in Table~\ref{tab:direct_st}. It is shown that the simplified ST-enhanced transfers can still outperform ASR-only ones in most of the cases, although the lack of ASR pre-training brings difficulties to ST model training.

\begin{table}[t]
    \centering
    \caption{Test WER for transfers without ASR pre-training}
    \begin{tabular}{c|cc|ccccc}
    \toprule
     & Vi & Ht & Pt & Zh-CN & Nl & Mn \\
     \midrule
    ASR & 54.5 & 59.4 & 39.5 & 39.2 & 43.0 & 67.7 \\
    ASR$\rightarrow$ST & 51.7 & 57.8 & 29.8 & 33.6 & 38.4 & 62.1 \\
    \midrule
    ST & 53.7 & 58.7 & 32.5 & 35.3 & \underline{44.1} & 67.3 \\
    \bottomrule
    \end{tabular}
    \label{tab:direct_st}
\end{table}



\begin{figure}[t]
    \centering
    \begin{tikzpicture}
        \begin{groupplot}[group style={group size= 2 by 1, horizontal sep=0.6cm, vertical sep=0 cm}, width=0.6\columnwidth, height = 0.49\columnwidth, legend cell align={left}, legend style={font=\scriptsize}]
            \nextgroupplot[title=Dutch, ymin=40.5, ymax=42.9, ylabel=Dev WER, y label style={yshift=-15pt}, xlabel=$n$, x label style={xshift=45pt}, xmin=0, xmax=6, legend style={legend pos=south east}, xminorticks=false, legend style={font=\scriptsize, at={(1.6,-0.62)},legend columns=3}]
            \addplot[blue, mark=square*] table [x index=0,y index=1, dashed, col sep=comma] {label_sampling_nl.csv};
            \addlegendentry{no filtering};
            \addplot[red, mark=*] table [x index=0,y index=2, dashed, col sep=comma] {label_sampling_nl.csv};
            \addlegendentry{10\% filtering};

            \nextgroupplot[title=Mongolian, ymin=56.85, ymax=59.1, xmin=0, xmax=6, legend style={legend pos=south east}, xminorticks=false, yticklabel pos=left]
            \addplot[blue, mark=square*] table [x index=0,y index=1, dashed, col sep=comma] {label_sampling_mn.csv};
            \addplot[red, mark=*] table [x index=0,y index=2, dashed, col sep=comma] {label_sampling_mn.csv};
        \end{groupplot}
    \end{tikzpicture}
    \caption{Dev WER for Dutch (highest MT resource) and Mongolian (lowest MT resource) ASR pre-trained with ST using N-best MT pseudo-labels (optionally with filtering).}
    \label{fig:label_sampling}
    \vspace{-5mm}
\end{figure}
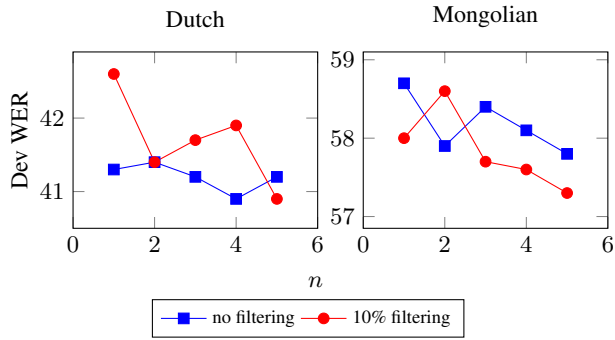
\begin{figure}[t]
    \centering
    \begin{tikzpicture}
        \begin{groupplot}[group style={group size= 2 by 1, horizontal sep=0.6cm, vertical sep=0.6 cm}, width=0.6\columnwidth, height = 0.49\columnwidth,
        legend cell align={left}, legend style={font=\scriptsize}]
            \nextgroupplot[title=Vietnamese, ymin=48, ymax=88, ylabel=Train accuracy, y label style={yshift=-15pt}, xlabel=Epoch, xlabel style={xshift=45pt}, xmin=-3, xmax=85, legend style={legend pos=south east}, xminorticks=false,
            legend style={font=\scriptsize, at={(1.7,-0.68)},legend columns=3},]
            \addplot[red, mark=*] table [x index=0,y index=3, dashed, col sep=comma] {acc_curve_vi.csv};
            \addlegendentry{w/ ST};
            \addplot[blue, mark=square*] table [x index=0,y index=2, dashed, col sep=comma] {acc_curve_vi.csv};
            \addlegendentry{w/o ST};
            \addplot[black, mark=triangle*] table [x index=0,y index=1, dashed, col sep=comma] {acc_curve_vi.csv};
            \addlegendentry{no transfer};
            
            \nextgroupplot[title=Mongolian, ymin=25, ymax=95, xmin=-3, xmax=38,
            xminorticks=false]
            \addplot[red, mark=*] table [x index=0,y index=3, dashed, col sep=comma] {acc_curve_mn.csv};
            \addplot[blue, mark=square*] table [x index=0,y index=2, dashed, col sep=comma] {acc_curve_mn.csv};
            \addplot[black, mark=triangle*] table [x index=0,y index=1, dashed, col sep=comma] {acc_curve_mn.csv};
            
        \end{groupplot}
    \end{tikzpicture}
    \caption{Training accuracy curve for Vietnamese (highest resource) and Mongolian (lowest resource).}
    \label{fig:acc_curve}
    \vspace{-5mm}
\end{figure}
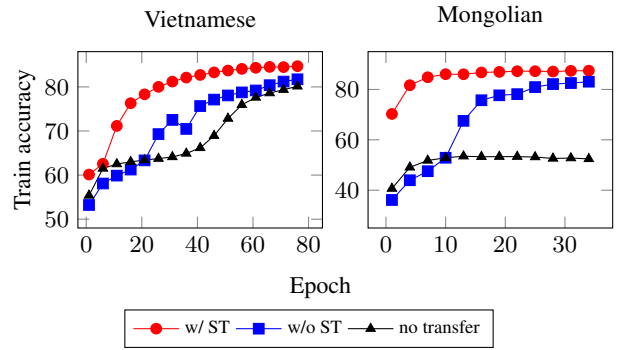

\section{Related Work}
End-to-end models, such as CTC models and attention-based encoder-decoder models, work well for low-resource ASR~\cite{rosenberg2017end}. It is known that multilingual training or pre-training with related languages improves low-resource end-to-end ASR significantly~\cite{dalmia2018sequence,yi2018adversarial,cho2018multilingual,zhou2018multilingual}. Meta learning methods~\cite{hsu2019meta} have recently been introduced to improve the efficiency of multilingual pre-training. Besides cross-lingual transfer learning, leveraging auxiliary data is another approach to improve low-resource ASR, for example, incorporating (synthetic) text translation data as additional inputs~\cite{hayashi2018back,anastasopoulos2018leveraging,Wiesner2019} or co-training with weakly supervised data~\cite{singh2019training} or text-to-speech (TTS) data~\cite{ren2019unsupervised}.

\section{Conclusions}

We show that cross-lingual (high-resource to lower-resource) transfer learning for end-to-end ASR can be improved by adding ST as an intermediate step. It makes transfer learning smoother in the two-step process and incorporates additional knowledge of the target language to improve model performance. It leverages only MT pseudo-labels but no expensive human labels to train ST and does not require high-resource MT training data. Currently, our approach is based on attention-based encoder-decoder architecture. Our future work includes extending this transfer learning approach to other end-to-end architectures, such as CTC and RNN Transducer.

\section{Acknowledgements}
We thank Ann Lee, Yatharth Saraf, Chunxi Liu and Anne Wu for helpful discussions.

\bibliographystyle{IEEEtran}
\bibliography{mybib}

\end{document}